\documentclass[]{aa}
\usepackage{graphicx}
\usepackage{natbib}
\usepackage{xtab,booktabs}

\usepackage{color} 
\usepackage{soul}

\usepackage{siunitx}

\begin{document}

\title{Pulsars with NenuFAR: backend and pipelines}


\author{L. Bondonneau
          \inst{1}
     \and
          J.-M. Grie{\ss}meier
          \inst{1,2}
     \and
          G. Theureau
          \inst{1,2,3}
     \and
          I. Cognard
          \inst{1,2}          
     \and
          M. Brionne
          \inst{1}          
     \and
          V. Kondratiev
          \inst{4,5}          
     \and
          A. Bilous
          \inst{4}          
     \and
          J. W. McKee 
          \inst{6}   
     \and
          P. Zarka
          \inst{14,2}  
     \and
          C. Viou
          \inst{2}              
     \and
          L. Guillemot
          \inst{1,2}
     \and
          S. Chen
          \inst{1}
     \and
         R. Main
          \inst{5}
     \and
         M. Pilia
          \inst{7}
     \and
         A. Possenti
          \inst{7,8}
     \and
         M. Serylak
          \inst{9,10}
     \and
         G. Shaifullah
          \inst{4}    
     \and
         C. Tiburzi
          \inst{4}    
     \and
         J. P. W. Verbiest
          \inst{11,5}    
     \and
         Z. Wu
          \inst{11}    
     \and
         O. Wucknitz
          \inst{5}    
     \and
          S. Yerin 
          \inst{12,13}       
     \and  
           C. Briand
          \inst{14}
     \and   
           B. Cecconi
          \inst{14,2}
     \and   
           S. Corbel
          \inst{2,15}
     \and   
           R. Dallier
          \inst{16,2}
     \and
           J. N. Girard
          \inst{15}
     \and   
           A. Loh
          \inst{14,2}  
     \and
           L. Martin
          \inst{16,2}
     \and
           C. Tasse
          \inst{17,2, 18}
 }

\date{Version of \today}

\institute{
        LPC2E - Universit\'{e} d'Orl\'{e}ans / CNRS, France
        \and
Station de Radioastronomie de Nan\c{c}ay, Observatoire de Paris - CNRS/INSU, USR 
  704 - Univ. Orl\'{e}ans, OSUC, Route de Souesmes, 18330 Nan\c{c}ay, France
        \and
        Laboratoire Univers et Th\'eories LUTh, Observatoire de Paris, CNRS/INSU, Universit\'e Paris
Diderot, 5 Place Jules Janssen, 92190 Meudon, France
        \and
        ASTRON, the Netherlands Institute for Radio Astronomy, Oude Hoogeveensedijk 4, Dwingeloo 7991 PD, the Netherlands
        \and
        Max-Planck-Institut f\"ur Radioastronomie, Auf dem H\"ugel 69, 53121 Bonn, Germany
        \and
        Canadian Institute for Theoretical Astrophysics, University of Toronto, 60 St. George Street, Toronto, ON M5S 3H8, Canada
        \and
        INAF-Osservatorio Astronomico di Cagliari, via della Scienza 5, 09047 Selargius, Italy
        \and
        Universit\'a di Cagliari, Dipartimento di Fisica, S.P. Monserrato-Sestu Km 0,700 - 09042 Monserrato, Italy
        \and
        South African Radio Astronomy Observatory, 2 Fir Street, Black River Park, Observatory 7925, South Africa
        \and
        Department of Physics and Astronomy, University of the Western Cape, Bellville, Cape Town 7535, South Africa
        \and
        Fakult\"at f\"ur Physik, Universit\"at Bielefeld, Postfach 100131, 33501 Bielefeld, Germany
        \and
        Department of Radio Astronomy Equipment and Methods of Observations, Institute of Radio Astronomy of NAS of Ukraine, Kharkiv, Ukraine
        \and
        Department of Astronomy and Space Computer Science, V. N. Karazin Kharkiv National University, Kharkiv, Ukraine
        \and
         LESIA, Observatoire de Paris, CNRS, PSL, SU/UP/UO, 92195 Meudon, France
        \and
        AIM, CEA, CNRS, Universit\'{e} de Paris, Universit\'{e} Paris-Saclay, F-91191 Gif-sur-Yvette, France
        \and
        SUBATECH, Institut Mines-Telecom Atlantique, CNRS/IN2P3, Universit\'{e} de Nantes, 44307 Nantes, France
        \and
        GEPI Observatoire de Paris, CNRS, PSL, SU/UP/UO, 92195 Meudon, France
        \and
        Centre for Radio Astronomy Techniques and Technologies, Department of Physics and Electronics, Rhodes University, Grahamstown 6140, South Africa
}

\date{Version of \today}
\abstract{
NenuFAR (New extension in Nan\c{c}ay upgrading LoFAR) is a new radio telescope developed and built on the site of the Nan\c{c}ay Radio Observatory. It is designed to observe the largely unexplored frequency window from 10 to 85\,MHz, offering a high sensitivity across its full bandwidth. NenuFAR has started its "early science" operation in July 2019, with 58\% of its final collecting area being available.
}
{Pulsars are one of the major topics for the scientific exploitation of this frequency range and represent an important challenge in terms of instrumentation. Designing instrumentation at these frequencies is complicated by the need to compensate for the effects of both the interstellar medium and the ionosphere on the observed signal. We have designed a dedicated backend and developed a complete pulsar observation and data analysis pipeline, that we describe in detail in the present paper,
together with first science results illustrating the diversity of pulsar observing modes available.
}
{
Our real-time pipeline LUPPI (Low frequency Ultimate Pulsar Processing Instrumentation) is able to cope with a high data rate and to provide real-time coherent de-dispersion down to the lowest frequencies reached by NenuFAR (10\,MHz). The full backend functionality is described, as well as the main pulsar observing modes (folded, single-pulse, waveform, and dynamic spectrum). This instrumentation allowed us to detect 172 pulsars in our first targeted search below 85\,MHz, including 10 millisecond pulsars (6 of which detected for the first time below 100 MHz).
}
{We also present some of the "early science" results of NenuFAR on pulsars: a high frequency resolution mapping of PSR B1919$+$21's emission profile and a detailed observation of single-pulse sub-structures from PSR~B0809$+$74 down to 16\,MHz, the high rate of giant-pulse emission from the Crab pulsar detected at 68.7\,MHz (43 events/min),  and the illustration of the very good timing performance of the instrumentation, allowing us to study dispersion measure variations in great detail. 
}
{}

\keywords{Pulsar, Low Frequency, Radioastronomy, Stars: neutron, Radio continuum: stars, Instrumentation: spectrographs, Interstellar medium: scintillation}

\titlerunning{Instrumentation for pulsar observations with NenuFAR}
\authorrunning{L. Bondonneau et al.}

\maketitle
\section{Introduction}



Pulsars are rapidly rotating, highly magnetised neutron stars, which emit collimated beams of radiation across the electromagnetic spectrum. 
Most of known pulsars have been discovered and studied in the radio band. 
According to the ATNF Pulsar Catalogue\footnote{http://www.atnf.csiro.au/people/pulsar/psrcat, V1.63} \citep{manchester_2005}, about 2800 pulsars have been found to date. Most of them have been discovered in the L-band (1--2\,GHz) with, e.g. the High Time Resolution Universe survey -- HTRU \citep{keith_2010,cameron_2020}, the Parkes Multi beam Pulsar Survey -- PMPS \citep{manchester_2001,lorimer_2015}, and the Pulsar Arecibo L-band Feed Array survey -- PALFA \citep{lazarus_2015,parent_2019}). A 
few hundred sources were unveiled by searches around 300--400\,MHz \citep{hessels_2008,lynch_2013}, 
and recently the low frequency domain was explored intensively with LOFAR, discovering 73 new pulsars 
in the window 119--151\,MHz with the LOTAAS survey \citep{sanidas_LOTAS_2019}. 

In the radio band, pulsars are known to have relatively steep spectra, with power-law index in the range of $-1$ to $-3$ \citep{malofeev_flux_2000, maron_pulsar_2000, bilous_lofar_2016}. Thus, despite the observational challenges caused mostly by  deleterious effects of propagation in the ionised interstellar medium, low-frequency pulsar surveys are promising. New low-frequency discoveries help sample the low luminosity end of the pulsar population and explore better the properties of the ionised interstellar medium, to which low frequencies are most sensitive. In addition, measuring pulsar flux densities at low radio frequencies helps constraining the so-called low-frequency turnover \citep{sieber_pulsar_1973, Malofeev1993, bilous_lofar_2020}, which is often present in pulsar spectra. Finally, the properties of radio emission at low frequencies become very dynamic, providing unique insight into conditions in the pulsar magnetosphere.

Several wideband pulsar studies have been performed recently at frequencies below 100\,MHz using pioneering radio telescopes such as LOFAR-LBA
\citep{hassall_wide-band_2012,pilia_wide-band_2016,kondratiev_lofar_2016,bilous_lofar_2020,bondonneau_2020}, UTR-2 \citep{zakharenko_detection_2013} and the LWA \citep{stovall_pulsar_2015}. LOFAR-LBA has a large fractional bandwidth, but highly non-uniform sensitivity, with antenna response peaking in the 45--75\,MHz range. UTR-2 operates at very low frequencies, 10--30\,MHz, but records a single linear polarisation. The LWA has proven to be very powerful, sharing the same antenna radiator design as NenuFAR \citep{Hicks2012} with a homogeneous frequency coverage across the band, but with only 256 antennas (to be compared with the present 1064 optimised NenuFAR antennas, that will eventually increase to 1824).

NenuFAR\footnote{\texttt{https://nenufar.obs-nancay.fr/en}} is a compact phased-array and interferometer, formed of hexagonal groups of 19 antennas called Mini-Arrays (MA). Its core counts today 56 MA = 1064 antennas. It will be extended to 80 MA = 1520 antennas by the end of 2020, and eventually reach 96 MA = 1824 antennas \citep{zarka_ursi_2020}.
Among current low-frequency telescopes, NenuFAR is well equipped to undertake an exhaustive census of the pulsar population in the low radio frequency band.

\section{Observing pulsars with NenuFAR}


Similarly to LOFAR's high frequency antennas \citep[HBA, see][]{van_haarlem_lofar_2013}, NenuFAR is designed to deal with the received radio signal on three different levels. 
While single antennas see the entire sky above them, analogically phased MAs of 19 antennas have a field of view of 8 to 69 degrees depending on the frequency (85--10\,MHz). This can be computed from the full width at half maximum (FWHM) $\frac{\lambda}{D}$, with $\lambda$ the wavelength and $D$ the MA diameter $\approx$25\,m. The full NenuFAR core with its 400 m diameter, allows to synthesise digital beams which are 0.5 to 4.2 degrees wide.
The geocentric coordinates of the phase centre are the same as for LOFAR international station FR606: (x,y,z) = (4324017.054, 165545.160, 4670271.072 m). The time reference comes from a rubidium clock source in Nan\c{c}ay.
The NenuFAR beamformer produces 768 simultaneous beamlets with 195\,kHz bandwidth at a time resolution of 5.12\,$\mu$s at any desired frequency in the band 10--85\,MHz and in any direction, within the analog beam of the phased MAs. 
This setup allows us in particular to cover the full bandwidth (75\,MHz) with two digital beams (768$\times$195\,kHz = 2$\times$75\,MHz), or to observe using four digital beams with a bandwidth of 37.5\,MHz each (768$\times$195\,kHz = 4$\times$37.5\,MHz).
By default, these four digital beams use all MAs; alternatively, different sub-arrays (with different sets of MAs) can be used for each of these four beams, allowing for a fly's eye mode with up to four totally independent directions on the sky, corresponding to four analogical beams.
For all cases, the full polarisation information of the signal is recorded in the form of four Stokes parameters.
More detail on NenuFAR are given in \citet{zarka_ursi_2020}. 
Flux and polarisation calibration are in progress and will be described in the full instrument paper to come (Zarka et al., in prep.).


\subsection{De-dispersion}

The main effect of the ionised interstellar medium is the dispersion of the radio waves, where the dispersive delay is proportional to the line-of-sight integrated electron content, the so-called dispersion measure (DM). Being also proportional to $\nu^{-2}$, where $\nu$ is the observing frequency, the dispersive delay has a strong impact on low frequency radio observations and requires dedicated instrumentation to process the signal and detect pulsed emission. 
The lowest frequency channel that can be coherently de-dispersed is directly related to the block size of the instrumentation ($\approx 10$ seconds) and the intra-channel dispersive delay corresponding to the DM of the observed pulsar, i.e.~12\,MHz for DM$\sim$10\,pc\,cm$^{-3}$, 25\,MHz for DM$\sim$100\,pc\,cm$^{-3}$, 
53\,MHz for DM$\sim$1000\,pc\,cm$^{-3}$.

\begin{figure}[tbp]
\centering
\includegraphics[height=8cm]{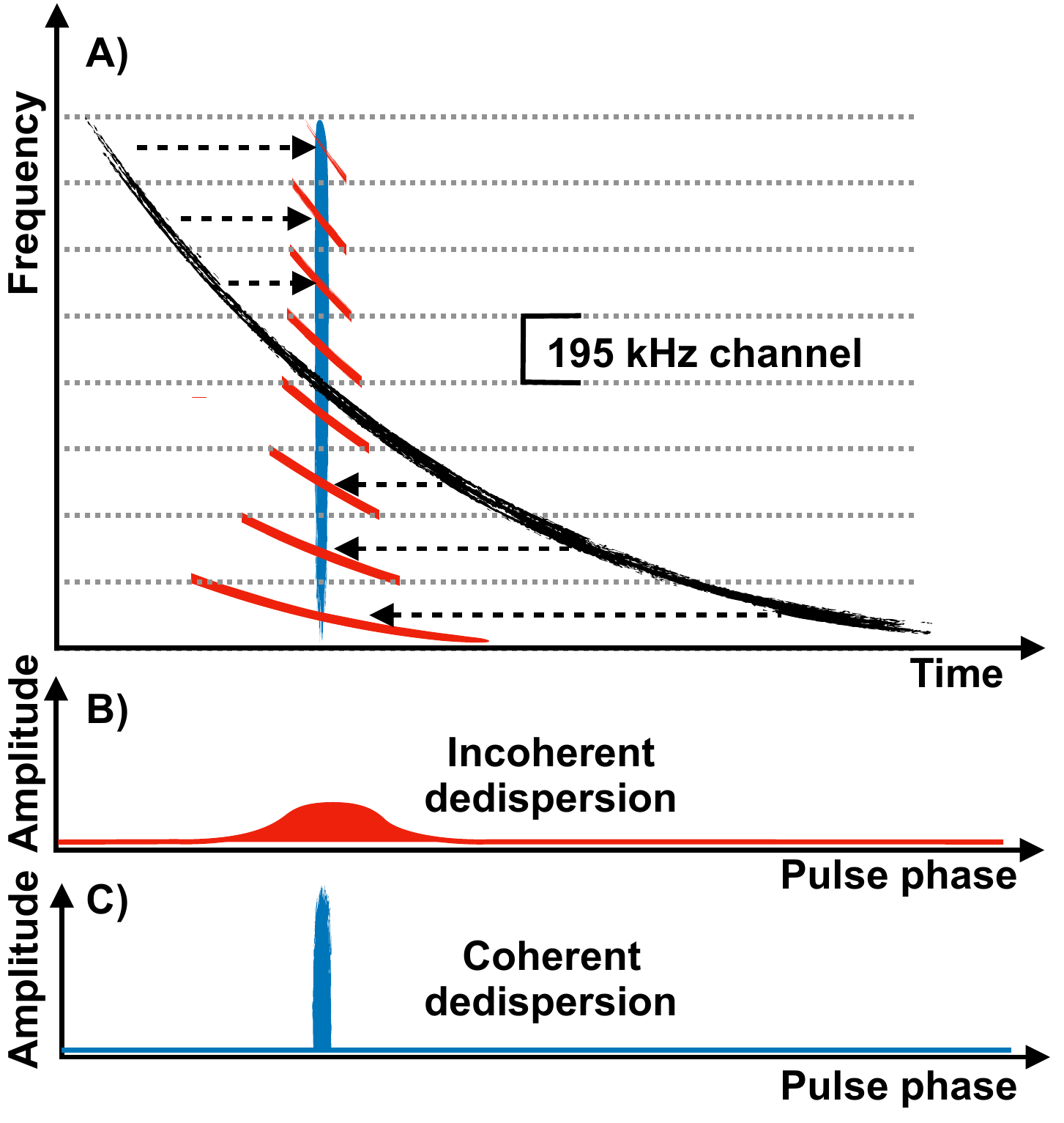}
\caption[Diagram comparing the two methods of de-dispersion]{\label{dedispersion_coherente} 
The top panel (A) shows a strongly dispersed pulsation from a pulsar in black, and the incoherently and coherently de-dispersed signal in red and blue, respectively. The channels are separated by grey dotted lines (195\,kHz wide for both LOFAR and NenuFAR). Panels (B) and (C) represent the de-dispersed and frequency integrated profiles using both methods. Amplitude scales are arbitrary but the areas below the red and blue curves should be similar.}
\label{tbp}
\end{figure}

There are two ways to correct for dispersive delays \citep[see e.g.][]{lorimer_handbook_2004, hankins_pulsar_1975}, as schematically shown in Fig.~\ref{tbp}.
Incoherent de-dispersion (in red) corrects for inter-channel delays only and, thus, requires a large number of narrow channels to minimise intra-channel smearing. It is generally applied off-line, once the signal has been recorded. 
Coherent de-dispersion (in blue) 
is designed to correct for intra-channels delay, managing the waveform signal 
by changing its phase in the Fourier domain. This process makes it possible to have wide channels while keeping a high time resolution. This method does not introduce dispersion smearing into the data in contrast with incoherent de-dispersion.
It is clear that coherent de-dispersion is necessary to preserve the integrity of the pulsar pulse profile, especially at low frequencies. To correct for the intra-channel dispersion at low frequency (10--85\,MHz) this method requires long Fourier transforms and a coherent dispersion code dimensioned for the corresponding buffer. As example the intra-channel dispersion delay of a signal with a DM value of 10\,pc\,cm$^{-3}$ is about 0.03~s at 85\,MHz, 0.13~sec at 50\,MHz, and 16.21~s at 10\,MHz.



\subsection{The UnDySPuTeD LUPPI}

The new real-time code LUPPI (Low frequency Ultimate Pulsar Processing Instrumentation, \citealt{bondonneau_pulsar_2018}) operating on NenuFAR
was adapted from NUPPI (Nan\c{c}ay Ultimate Pulsar Processing Instrument, \citealt{desvignes_coherent_2014}), the Graphical Processing Unit (GPU) cluster for the Nan\c{c}ay Radio Telescope (NRT) observations since August 2011. NUPPI is designed to handle dispersive delays in 4\,MHz channels in L- or S- bands for a total instantaneous bandwidth of 512\,MHz. 
The NUPPI software was derived from GUPPI (Green Bank Ultimate Pulsar Processing Instrument, \citealt{demorest_data_2014} and \citealt{duplain_launching_2008}).

LUPPI runs on two composite machines named UnDySPuTeD (Unified Dynamic Spectrum Pulsar and Time Domain receiver), each hosting two Intel Xeon E5-2620 CPUs (Central Processing Units) and two Nvidia GeForce GTX 1080 GPUs.

By using together the two UnDySPuTeD machines, LUPPI has at its disposal a total of 32 CPU cores, 512\,GB of error-correcting code RAM memory and 32\,BG of GPU memory. Because of the observing frequencies below 100\,MHz, LUPPI has to deal with additional difficulties compared to GUPPI and NUPPI.

The huge intra-channel dispersion imposes extremely long Fourier transforms (typically more than 20 seconds, i.e. 200 times longer than for similar computations at 1.4\,GHz) and leads to memory management difficulties (both RAM and GPU) in the NenuFAR framework. These difficulties have been mitigated by modifying the size of the buffers as well as the variables that contain it in the source code. These modifications enable the correction of intra-channel dispersive delays up to 10.737\,s.

\begin{figure}
\centering
\includegraphics[height=9cm]{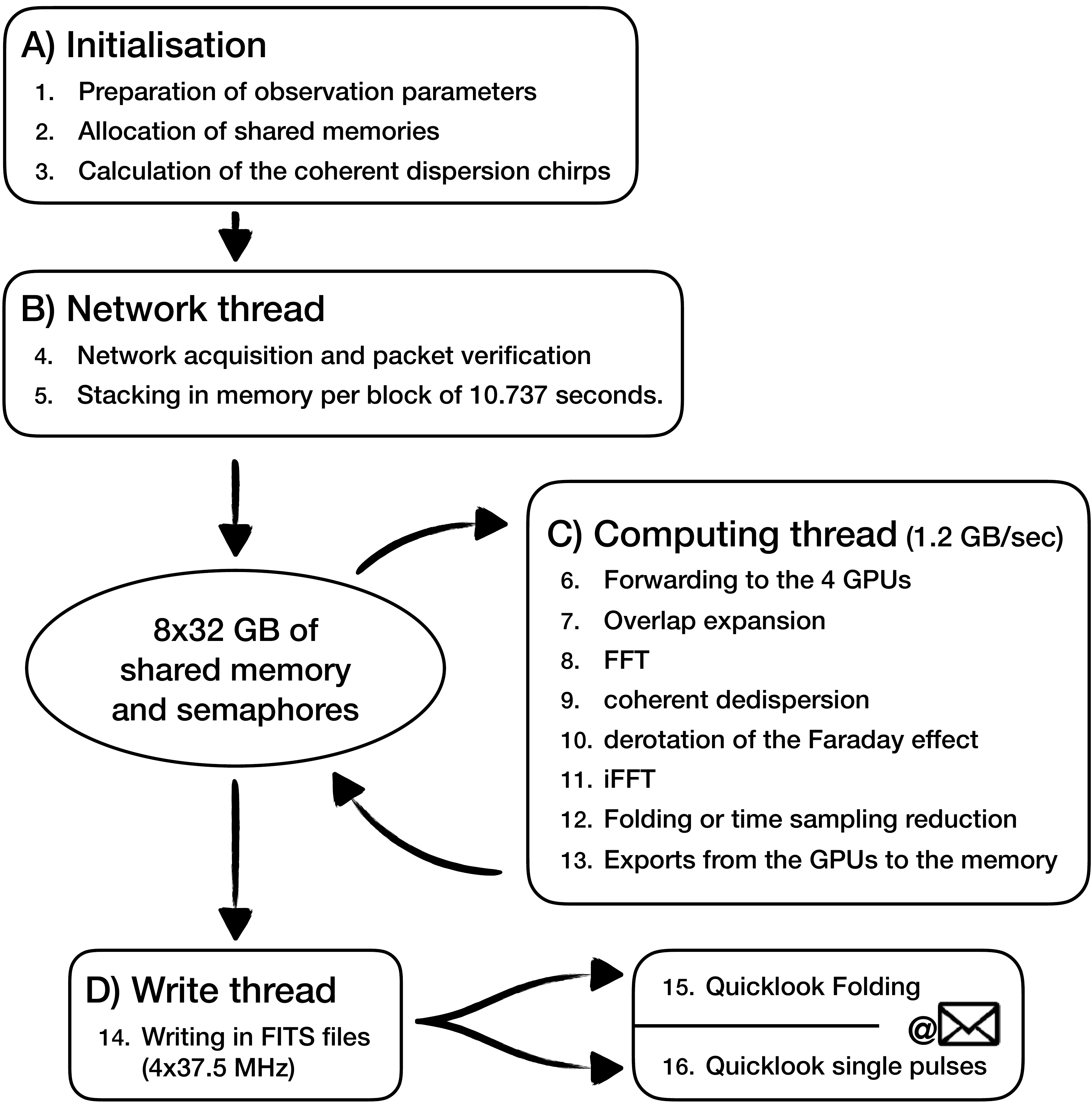}
\caption[LUPPI]{Operational scheme of the LUPPI pipeline including its main elements: Initialisation (A) of the observation parameters, the data processing cycle organized in threads around a shared memory and communicating through semaphores and post-processing of the observation with production of ``quicklooks'' plots.
The network thread (B) catches the packets send by the beamformer, 
the computing thread (C) processes the data in the Fourier domain on GPUs and the write thread (D) exports data products to disks.}
\label{LUPPI}
\end{figure}

LUPPI captures the 768 beamlets produced by the NenuFAR beamformer (described in~\citealt{zarka_ursi_2018, viou_2019})
as a series of network packets.
The two complex-voltage values (one per polarisation) at each time step are used to compute the four Stokes parameters of the signal.

The operational scheme of LUPPI is shown in Fig.~\ref{LUPPI}. The pipeline is divided into four real-time elements and a post-processing analysis:

A) The initialisation phase prepares the parameters of the observation (DM, channels, source name, etc.) and allocates a total of 128~GB\footnote{\texttt{buffers size:}  4 (4$\times$37.5\,MHz) $\times$ 2 (input and output buffer) $\times$ 4 (number of blocks per buffer) $\times$ 2 (dual linear polarisation) $\times$ 2 (2 byte complex) $\times$ $2^{22}$ (number of points per Fast Fourier Transform (FFT) $\times$ 192 (number of channels) $\simeq$ 8$\times$12.88~GB.  The buffer size is rounded up to 8$\times$32~GB to be compatible with a future 16-bit mode.}
of shared memory. We use shared memory to implement ring buffers\footnote{\texttt{The ring buffer:} a structure for storing data in memory where the last block is topologically adjacent to the first one, giving it its ring characteristic.} in order to allow threads to exchange data and process in parallel. Furthermore it is during this phase that the coherent dispersion chirps\footnote{\texttt{The chirp:} transfer function for the phase changes of the coherent dispersion in the Fourier domain. As the chirp is frequency dependent, there are as many chirps as there are channels per single DM.} are calculated.

B) The network thread catches UDP (User Datagram Protocol) packets sent by the beamformer and reads the packet index to look for any missing packets and takes it into consideration to preserve the linearity of the time series.
The packets are then stored in the shared memory and a semaphore is used to unleash the computation thread. 

C) The computation thread is dedicated to feed GPUs with data and get the results back. At this stage, the data blocks are sent to four GPUs (one processing unit per band of 37.5\,MHz).

First, a fixed overlap\footnote{\texttt{The overlap:} a duplication of the ends of the data series to take into account the differences in dispersive delays between the top and bottom of the lowest frequency channel during the FFT. 
}
is added 
to absorb the intra-channel dispersive delay.
%
For a given GPU hardware, the value of the overlap results as a compromise between computing efficiency (which is better when the overlap is small) and the possibility to reach low frequencies (for a given DM value, a larger overlap allows to process data at a lower frequency).
With our GPUs being limited to 8\,BG of memory, 
we chose an overlap of 50\% of the total block size, i.e.~10.737 seconds.
With this, we can de-disperse frequency channels
of frequency
>12\,MHz for a DM of $\sim$10\,pc\,cm$^{-3}$, >25\,MHz for a DM of $\sim$100\,pc\,cm$^{-3}$, 
>53\,MHz for a DM of $\sim$1000\,pc\,cm$^{-3}$.

Then, the complex voltages are transformed to the Fourier domain using the NVIDIA CUDA Fast Fourier Transform library (cuFFT \texttt{https://docs.nvidia.com/cuda/cufft}) in order to multiply the signal by the transfer function corresponding to the coherent dispersion chirp. One of the unique features of this instrumentation is the availability of a Faraday coherent de-rotation allowing to correct for the rotation of the linear polarisation within each channel (Bondonneau et al., in prep).

Furthermore, the data are transformed back to the time domain with a inverse Fast Fourier Transform (iFFT) and folded at the period of the observed pulsar or integrated in time depending on the observing mode. Finally, the result is pushed into the output buffer and a semaphore is activated to launch the write thread.

D) The write thread 
writes the data product out from the output buffer into a RAID 0 (Redundant Array of Independent Disks using data striping) storage unit.

The last step is the post-processing to verify that the observation output products are present on the server and to create diagnostic plots (``quicklooks'', see \S\ref{section_quicklook}). The quicklook plots are sent by email to the observer within minutes. If data files are missing on the server a warning email is sent to the observer. The source code of LUPPI is available on the 
Github\footnote{\texttt{https://github.com/louisbondonneau/LUPPI}}.

\subsection{Currently available observing modes}

NenuFAR and LUPPI currently offer several observing modes for pulsars. We describe the main ones below.

\subsubsection{Folded mode}
\label{section_folded}

Time series are coherently de-dispersed in every 195\,kHz channel and folded at the apparent period of a pulsar using an up-to-date ephemeris. The default sub-integration time is 10.737\,s. Data are stored in PSRFITS\footnote{\texttt{https://www.atnf.csiro.au/research/pulsar/psrfits\_definition/Psrfits.html}} format \citep{hotan_psrchive_2004} and a quicklook plotting program displays the main features of the observation after removal of radio frequency interference (RFI). This is the main observing mode. It is used to characterise the pulse profile and polarisation as a function of frequency, analyse the spectral energy distribution, and study long-term variations, e.g.~related to DM fluctuations or intrinsic behaviour in the pulsar magnetosphere.\\

\subsubsection{Single-pulse mode}
\label{section_singlepulse}

In this mode a time series is de-dispersed
and integrated in time by a chosen factor (128 by default) to enable the analysis of  
pulse-to-pulse variations. Data are stored in PSRFITS format and a quicklook program displays the main features of the observation after RFI removal (using the \texttt{rfifind} software from \texttt{PRESTO}\footnote{\texttt{PRESTO}: a toolbox developed to search for single pulses and periodic pulsations in pulsar observations  \citep[see][]{ransom_PRESTO_2001}.}, \citealt{ransom_PRESTO_2001}).
This mode is used to observe bright pulsars that show interesting features in their emission beam, such as drifting subpulses, mode changing, or anomalously intense pulses. It will also be used to search for pulsed emission when the DM is higher than 100\,pc\,cm$^{-3}$ and the scattering is important.
Moreover, single pulse data can easily be converted into folded data.\\

\subsubsection{Waveform mode}
\label{section_waveform}

This mode consists in writing the raw complex-voltage data to disk, bypassing the computing thread on Fig.~\ref{LUPPI}. It keeps the full time, frequency and polarisation information from the original raw signal (thus a very large data rate). This mode can be used to perform non-standard analyses, such as scattered giant pulses searches, a census of the Globular Clusters pulsar population, or interstellar and interbinary scintillation studies. Additionally, waveform data can be converted into folded or single pulse data by coherent de-dispersion and integration.\\
The two UnDySPuTeD machines also provide LUPPI with additional, independent modes described below:

\subsubsection{Dynamic spectrum mode}
\label{section_dynspec}

\begin{figure}
\centering
\includegraphics[width=9cm]{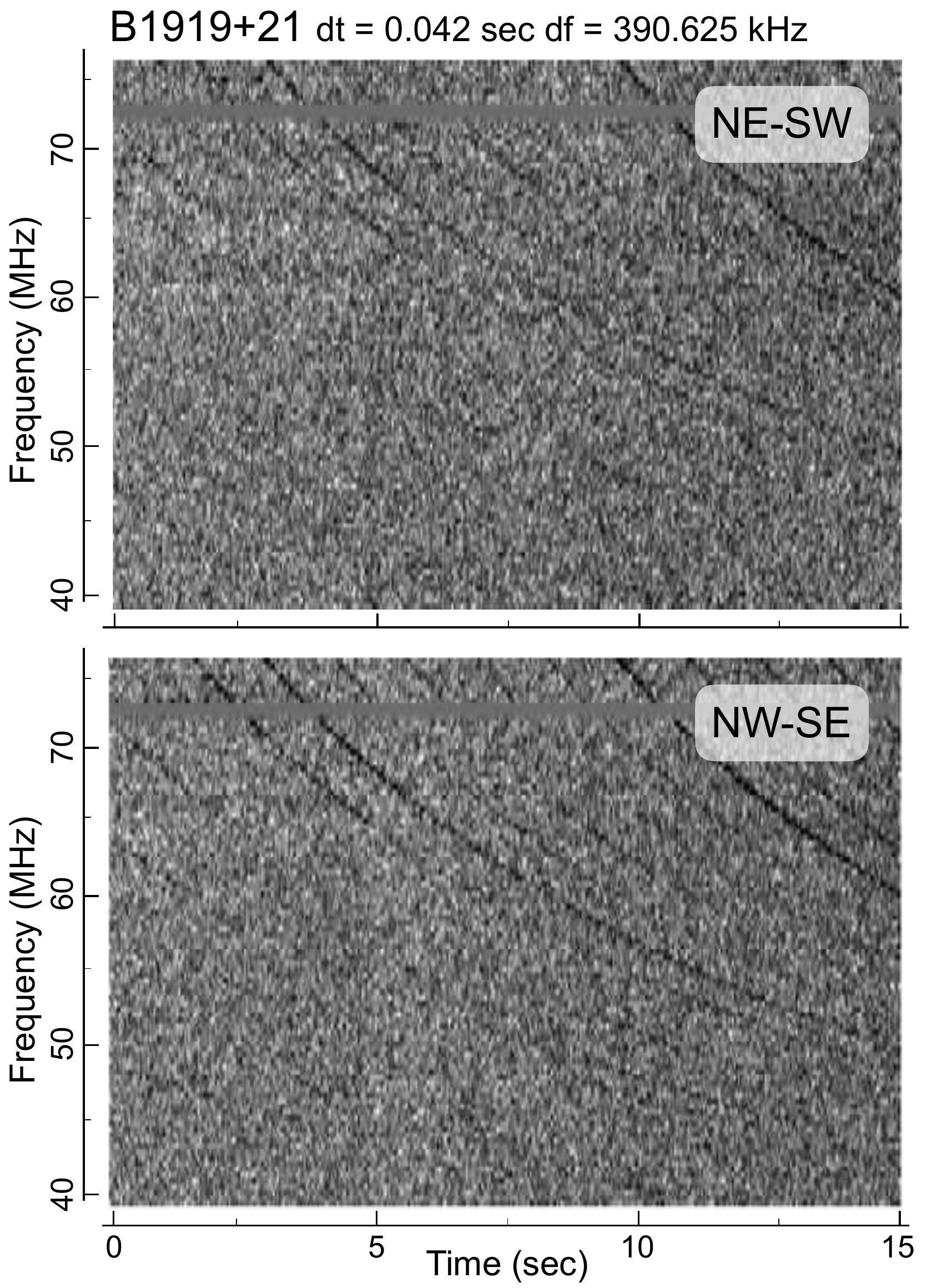}
\caption[B1919_dynspec]{Dynamic spectrum of a 15~min observation pointing on PSR~B1919$+$21. The top and bottom panels provide the linear feed polarisation from the NE-SW and NW-SE dipoles corresponding to XX and YY, respectively. These dynamic spectra were recorded with time and frequency resolutions of 0.01~s and 1.5\,kHz, and then integrated to 0.04~s and 390\,kHz for this display.}
\label{B1919_dynspec}
\end{figure}

\begin{figure*}[!]
\centering
\includegraphics[width=18cm]{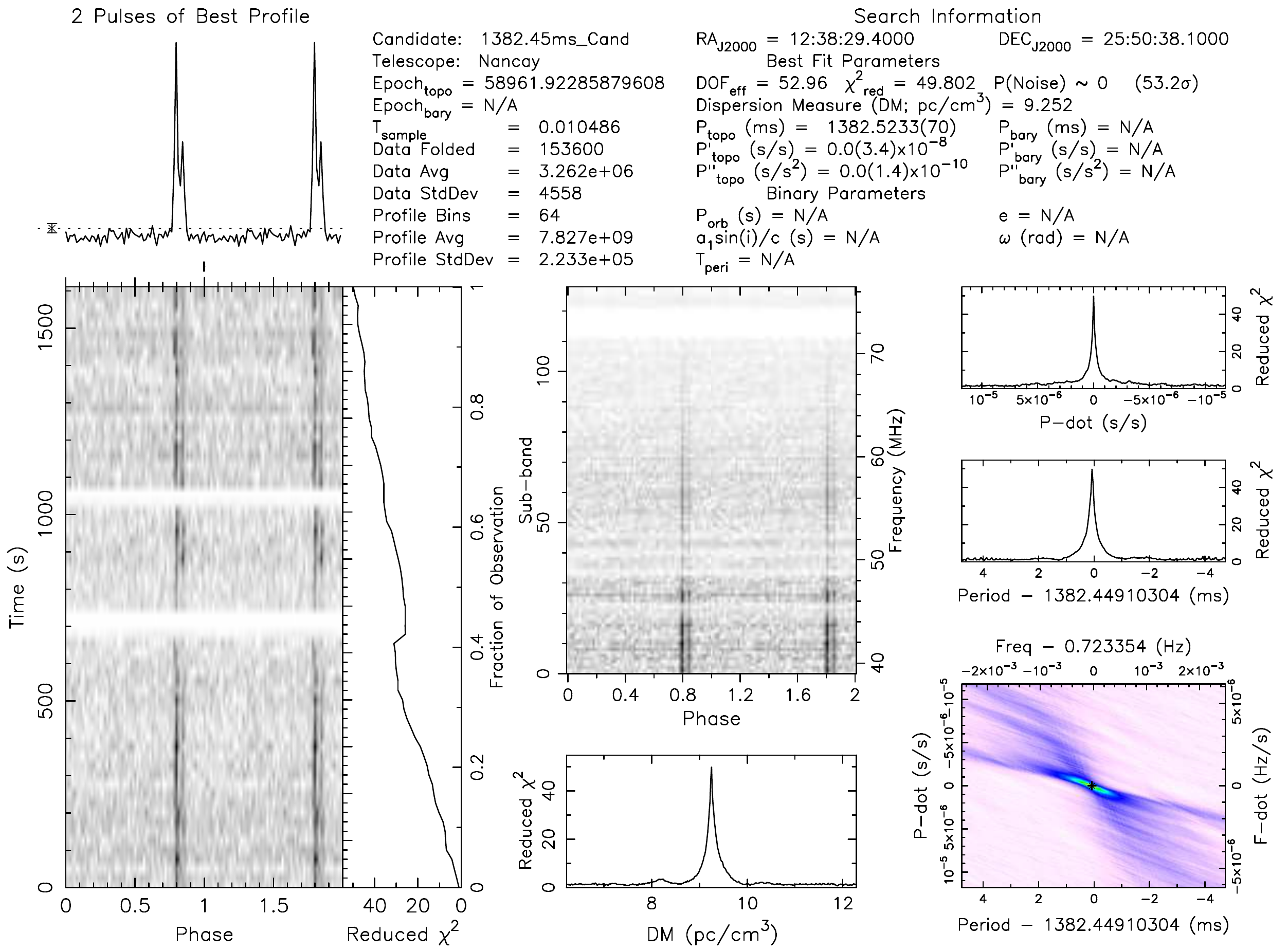}
\caption[B1237+25]{Detection plot of PSR~B1237$+$25, the first candidate provided from the \texttt{PRESTO} analysis of a 30-min observation. The blind search was conducted over DMs between 1.0 and 60.0 with a step of 0.01\,pc\,cm$^{-3}$.
Top-left panel: detection profile over two periods for the best DM and pulsar spin period, P0. Bottom-left panel: time vs. phase diagram for the best DM and P0. Right part of the bottom-left panel is the reduced $\chi^2$ obtained from the stacked profile relative to the corresponding stacked median and variance at each time. Middle panel: spectrum vs. phase diagram for the best P0 and DM. Bottom-middle panel: reduced $\chi^2$ obtained for each trial DM. Bottom-right panel: reduced $\chi^2$ obtained for each trial (P0, P0 time derivative $\dot{P}$) at the best DM. Top-right panels: reduced $\chi^2$ relative to $\dot{P}$ for the best P0 and relative to P0 for the best $\dot{P}$.}
\label{B1237}
\end{figure*}

The 195\,kHz wide beamlets sent by the beamformer are GPU-processed to provide finer frequency resolution and time integration.
The frequency resolution can be chosen between 0.10 and 12.20\,kHz (corresponding to FFT lengths from 2048 to 16 samples).
The time resolution (corresponding to a number of integrated FFTs restricted to be a power of 2, and being at least 4) can be chosen between 0.30 and 84.0\,ms (depending on the frequency resolution).
An apodisation function (windowing) can be applied among the few ones proposed (such as Hann, Hamming, or others), the default being Hamming.
The resulting dynamic spectra are recorded in a very simple binary format.
This dynamic spectrum mode is used for solar or Jupiter observations, exoplanet searches, and in the context of a pulsar blind survey or any radio transient search. This mode can be used to observe pulsars without de-dispersion, for instance the bright pulsar B1919$+$21 (Fig.\ref{B1919_dynspec}). In addition, it supports very fine channels (e.g. 1.5\,kHz) enabling searches over wider DM ranges without coherent de-dispersion (see a detection example in Fig.~\ref{B1237}).\\

\subsubsection{Compressed UDP waveform mode}
\label{section_udp_waveform}

For a small number of projects we use a second waveform mode, in which UDP packets are directly dumped to disk when they arrive on the network socket.
In this mode, we use real-time lossless compression with \texttt{zstd}\footnote{\url{https://github.com/facebook/zstd}}, which allows us to reduce the data volume by a factor of $\sim$2-3. Files written in this mode do not contain any header information. This mode is used, for example, for low-frequency Very Long Baseline Interferometry (VLBI) with international LOFAR stations on baselines of $\sim$1000 km.\\

\subsection{The post-processing pipeline}

In order to analyse observations we have developed two post-processing pipelines for folded and single-pulse observations. The role of these pipelines is to ensure that the transfer between the real-time processing machine and the data storage is successful and to redirect the data into the different branches of processing according to their mode of observation: waveforms, folded or single-pulses.

\subsubsection{Waveform mode}
The waveform mode corresponding to the raw data requires large processing and storage resources (540\,BG per hour 
for a bandwidth of 37.5\,MHz), consequently it is not handled by the quicklook pipeline even if the transfer remains automatic.

\subsubsection{Folded mode}
\label{RFImitigation}
The post-processing pipeline for the folded observations is based on Nenuplot.py\footnote{\texttt{Nenuplot.py:}
a completely automatised pipeline for folded observations. Based on the \texttt{PSRCHIVE} library \citep{hotan_psrchive_2004} this code is able to rapidly mitigate RFIs in the observation (using a modified version of the  \texttt{CoastGuard} software, \citealp[][\texttt{https://github.com/plazar/coast\_guard}]{lazarus_prospects_2016}) and create a quicklook. The source code is available on Github \texttt{https://github.com/louisbondonneau/NenuPlot}}. Based on the \texttt{PSRCHIVE}\footnote{\texttt{PSRCHIVE}: a library for the analysis of pulsar astronomical data. It implements an extensive range of algorithms for use in pulsar timing, scintillation studies, polarimetric calibration, single-pulse work, RFI mitigation, etc. The software is described in \citet{hotan_psrchive_2004}.}library. This program is fed with the folded FITS files to perform the RFI mitigation and get out a quicklook in PNG format as described in the next section. Finally, the quicklook is automatically sent by e-mail to the observer.

\subsubsection{Single-pulse mode}
Single pulse observations are converted from 32 bits to 8 bits and normalised in time and frequency domains with a fits file converter\footnote{\texttt{ChangeFormat\_rescale.py}: This code is designed to convert 32-bit PSRFITS data into 8-bit in data while maintaining the signal fidelity. The signal is transformed to 8-bit by decomposing it in data, scale and offset (signal$=$(data$\times$scale)$+$offset) which allows us to maintain the dynamic range of the signal. The source code is available on Github \texttt{https://github.com/louisbondonneau/psrfits\_search}}. Furthermore, the blocksize of a PSRFITS file is reduced with \texttt{RescaleTime.py}\footnote{\texttt{RescaleTime.py:} The objective of this code is to reduce the blocksize in the PSRFITS file. The source code is available on Github \texttt{https://github.com/louisbondonneau/psrfits\_search}} in order to allow more accurate RFI mitigation with \texttt{rfifind}. Then, the PSRFITS file is analysed with \texttt{PRESTO} tools using a fine DM grid adapted to low frequencies to search for intense pulses in time and DM. \texttt{PRESTO} produces a quicklook analog to that shown in Fig.~\ref{crabe} which is sent by e-mail to the observer. Moreover, we use \texttt{dspsr}\footnote{\texttt{DSPSR} \citep{van_straten_dspsr_2011} is an Open Source library. It is available on Github \texttt{https://github.com/demorest/dspsr}.} to transform single-pulse 8-bit PSRFITS files into folded and de-dispersed PSRFITS files allowing us to use \texttt{Nenuplot.py} and produce a quicklook (see in \S\ref{section_quicklook}) and send it to the observer with the quicklook of \texttt{PRESTO}.

\subsubsection{The quicklook}\label{section_quicklook}

\begin{figure*}[!]
\begin{center}
\includegraphics[scale=0.90, trim=0cm 1cm 0cm 0cm]{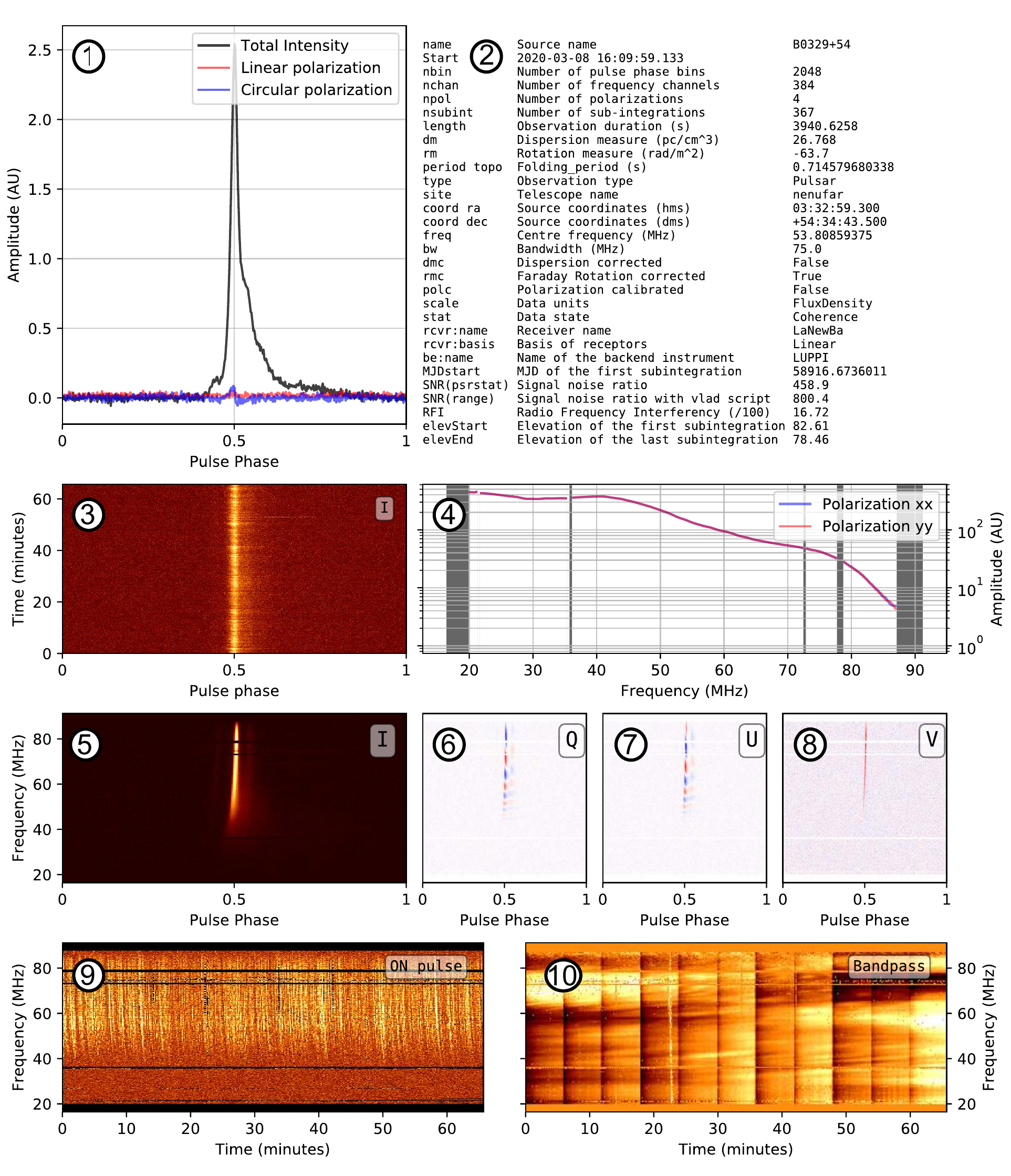}
\end{center}
\caption
{Quicklook of an observation of PSR~B0329$+$54 with NenuFAR in folded mode. Panel 1: profile of the pulsar in total intensity (black), linear polarisation ($\sqrt{(Q^2+U^2)}$, red), and circular polarisation (V, blue). Panel 2: metadata of the observation. Panel 3: time variation of the frequency-integrated profile. Panel 4: bandpass (spectral response of the telescope) for the linear feed polarisation parameters XX and YY (in blue and red, respectively). Panels 5, 6, 7, and 8 show the frequency-resolved profiles for the Stokes parameters I, Q, U, and V. Panel 9: dynamic spectrum of the amplitude of the  pulsation (``ON-pulse region''). Panel 10: dynamic spectrum of the bandpass (summmed on ``OFF-pulse'' regions).}

\label{quicklook}
\end{figure*}

The creation of a quicklook is a crucial part of the observation because it makes it possible to monitor the operation of the pipeline as well as the quality of the observation with respect to, e.g., RFI. Its aim is to allow quick validation of an observation and to reduce the response time in case of problem.\\
This python software is based on the \texttt{PSRCHIVE} library and it is designed to be a complete pipeline compatible with all radio telescopes generating folded PSRFITS files. A quicklook of an observation of PSR~B0329$+$54 is shown in Fig.~\ref{quicklook}. The quicklook plot presents the observation data in all its dimensions: time, phase, frequency, and polarisation in various panels. Panel 1 
shows the pulse profile with the total intensity (in black), the linear polarisation (in red), and the circular polarisation (in blue). The profile allows us to determine if the observed source is detected or not. Panel 2 shows the metadata of the observation 
including the source name, DM, rotation measure (RM), P0, observation duration, bandwidth, signal-to-noise ratio (SNR), percentage of RFI, elevation of the source, etc. Panel 3 
shows the temporal variation of the frequency-integrated profile. This plot is useful to highlight changes in pulsar period with respect to the reference ephemeris used for folding the observation or a time variation of the pulse profile.
Panel 4 is the time-integrated bandpass (i.e. the spectral response of the telescope) for the polarisation parameters XX and YY (in blue and red, respectively). 
Channels flagged because of the presence of RFI are highlighted in grey in this panel. Panel 5 
shows the time-integrated profile as a function of frequency for Stokes I (total intensity).
This plot is important because at low frequency we are particularly sensitive to changes in DM.
Panels 6, 7, and 8 show the frequency resolved profiles for the Stokes parameters Q, U, and V. The colour corresponds to the sign of the parameter (red for positive values and blue for negative). These plots are helpful to characterise the RM used to correct for the Faraday rotation between channels. In the example of  Fig.~\ref{quicklook}
there is a frequency dependent structure in the parameters Q and U. This is the signature of an imperfectly corrected Faraday effect caused by ionospheric variations (i.e.~the total RM is different from the value in the reference ephemeris used during the observation). Panel 9 shows the dynamic spectrum of the amplitude of the pulsation (recorded in the ``ON-pulse'' region), which can be used to quantify mitigation and data quality.
Panel 10 is the dynamic spectrum of the bandpass 
(spectral response from the ``OFF-pulse'' regions) that can be used to assess the quality of the observation. 
The scintillation of the source is visible on panel 9, while panel 10 shows a 6-min periodic drop of the bandpass (gain) that result from the analog pointing of the instrument and can be corrected in post-processing.\\

\subsection{Data archive}

Currently, NenuFAR pulsar data are transferred to the Nancay Data Center (CDN) for temporary storage and distribution within the community.
At the CDN, the data are stored on a RAID 6 storage cluster. 
Pulsar data recorded in folded mode (\S \ref{section_folded}), single-pulse mode  (\S \ref{section_singlepulse}), and dynamic spectrum  (\S \ref{section_dynspec}) mode will be kept at the CDN, whereas waveform data  (\S \ref{section_waveform}) and data recorded in UDP waveform mode (\S \ref{section_udp_waveform}) are usually only kept temporarily and are removed after processing.
A dedicated NenuFAR Data Centre is currently being commissioned, it will have open access user interfaces to data products.

\section{Commissioning and first early science results}

Early Science phase of the instrument will last until the end of 2021. During this time we will be
conducting a set of 
observational programs with NenuFAR, under the umbrella of the ``Pulsars Key Science Project'' (ES03).
Up to now, all our observations were made with a limited
version of the full instrument, i.e. the 56-MA first step configuration that started operating with a stable setup in February 2019 \citep{zarka_ursi_2020}. A 80-MA configuration will be operational by the end of 2020.

\subsection{First NenuFAR pulsar catalogue}

A first targeted survey has already  allowed us to observe 650 pulsars with DM $<100$\,pc\,cm$^{-3}$ above \SI{-20}{\degree} of declination, of which about 26\% were detected (Bondonneau et al., in prep). Where the non-detections are mainly due to the amplitude of the scattering at theses frequencies.
These observations are an important complement to recent studies of the pulsar population observable below 85\,MHz such as \citet{bilous_lofar_2020} and \citet{bondonneau_2020}.


This first NenuFAR catalogue, combined with the first measurements of flux density and polarisation, 
is now used as a basis for a set of follow-up programs, 
e.g. the study of pulse-to-pulse variations with time and frequency, the effects of scintillation due to the interstellar medium, the characterisation of the pulsar population and the eclipses of certain systems. A set of 41 bright pulsars is tracked monthly to study the long-term variations of their average profiles, to characterise the DM evolution, and to study their spectral energy distribution in detail. 
Finally, a blind survey of the northern sky visible from Nan\c{c}ay started in Summer 2020 and will provide a more complete census of the local pulsar population at frequencies below 100\,MHz.

\begin{figure*}[ht]
\centering
\includegraphics[height=13cm, trim=0cm 0cm 0cm 0cm]{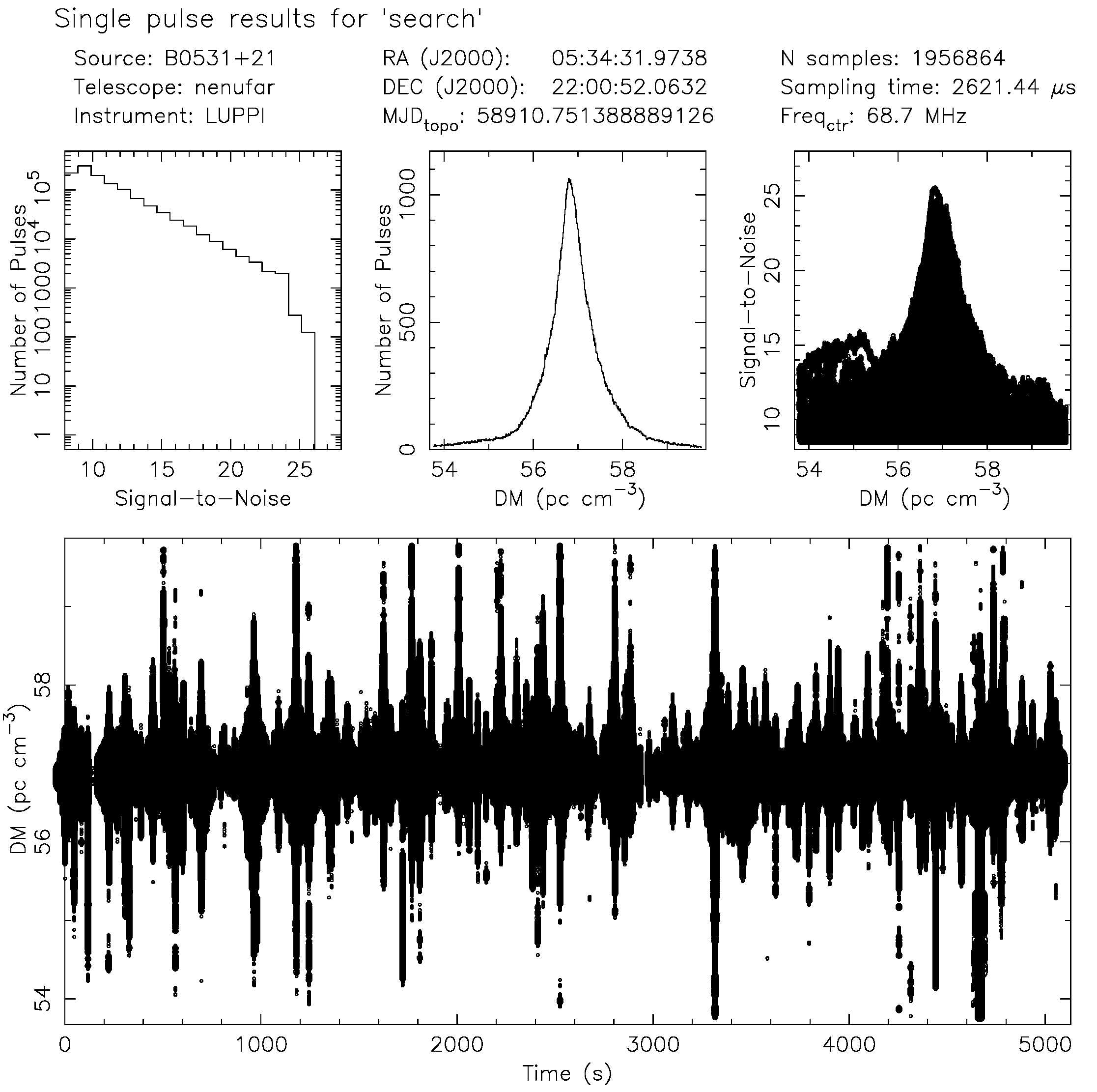}
\caption[Crab pulsar giant pulses]{Giant pulses from the Crab pulsar (PSR~B0531$+$21) detected as a function of time and trial DM values. Bottom panel: detection time for all trial DMs. Due to the pulse width and scattering the same pulse can be detected over a wide range of DMs and count as multiple detection in top panels. Top-left and top-middle panels show histograms of giant pulses' SNRs and trial DMs. Top-right panel: histogram of giant pulses' SNRs vs. trial DM values.
}
\label{crabe}
\end{figure*}

\subsection{Details of the emission profile of PSR~B1919$+$21}

\begin{figure}
\centering
\includegraphics[height=8.5cm]{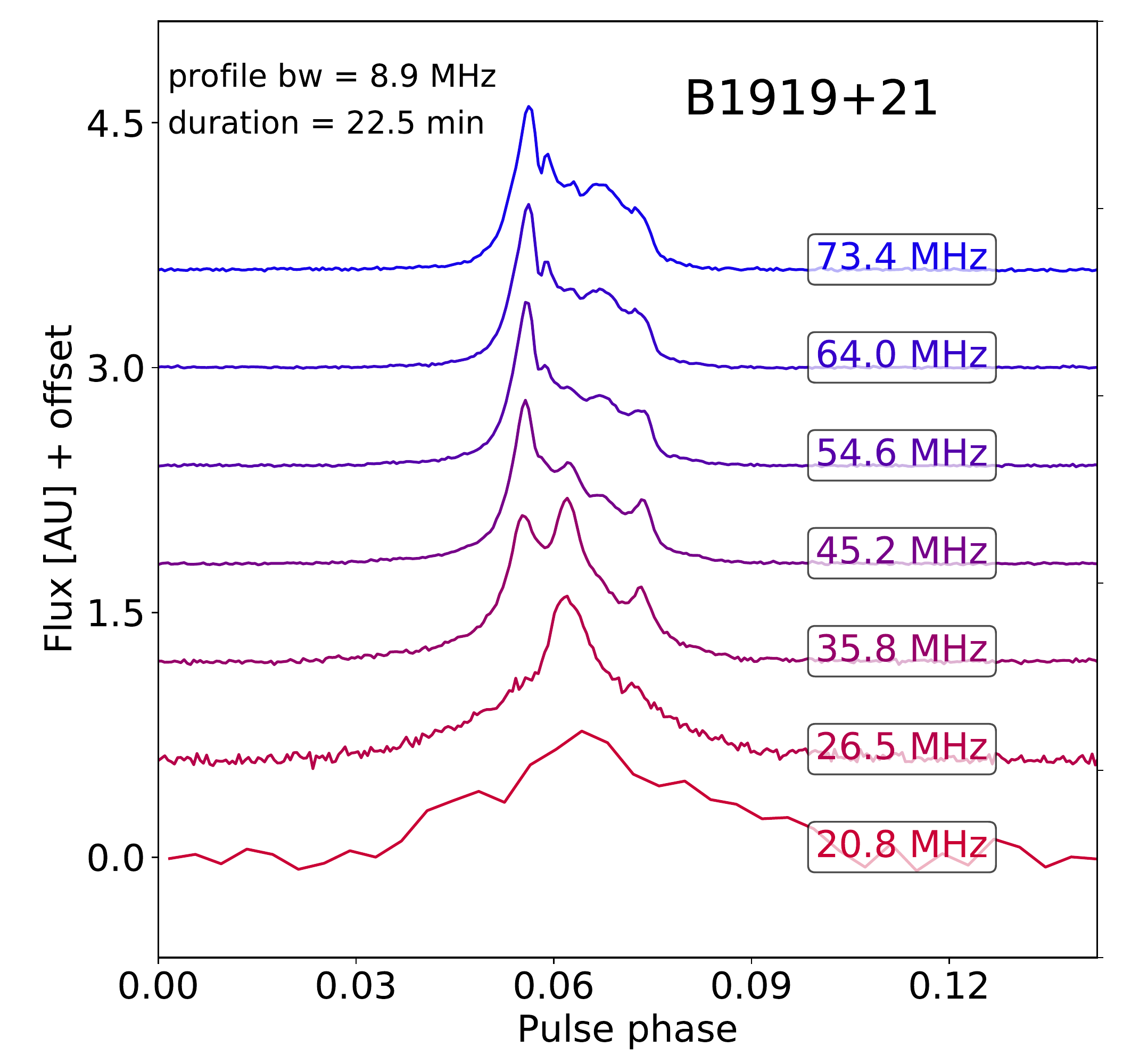}
\caption[B1919+21]{Profile variation of PSR~B1919$+$21 with frequency. Each line corresponds to an integrated and normalised profile on a frequency band of 8.9\,MHz. The total integration time is 22.5 min.}
\label{B1919}
\end{figure}

The high sensitivity of NenuFAR and its stable gain across the band make it ideally suited to study average profiles in detail. Figure~\ref{B1919} shows the evolution of the observed profile for the historical pulsar B1919$+$21, discovered by Jocelyn Bell in 1967 \citep{hewish_observation_1968}. Using NenuFAR, this pulsar has been detected with a high SNR ($>2000$) down to 20\,MHz. Obtaining good-quality data at such low frequencies is important for modelling profile evolution and constraining emission heights. About three dozen sources have been observed with such frequency resolution as in Fig.~\ref{B1919} in integrations of only one hour \citep{bondonneau_thesis_2019, bondonneau_2020}.

\subsection{Crab pulsar giant pulses}

\begin{figure}[ht!]
\centering
\includegraphics[height=9cm, trim=0cm 0cm 0cm 0cm]{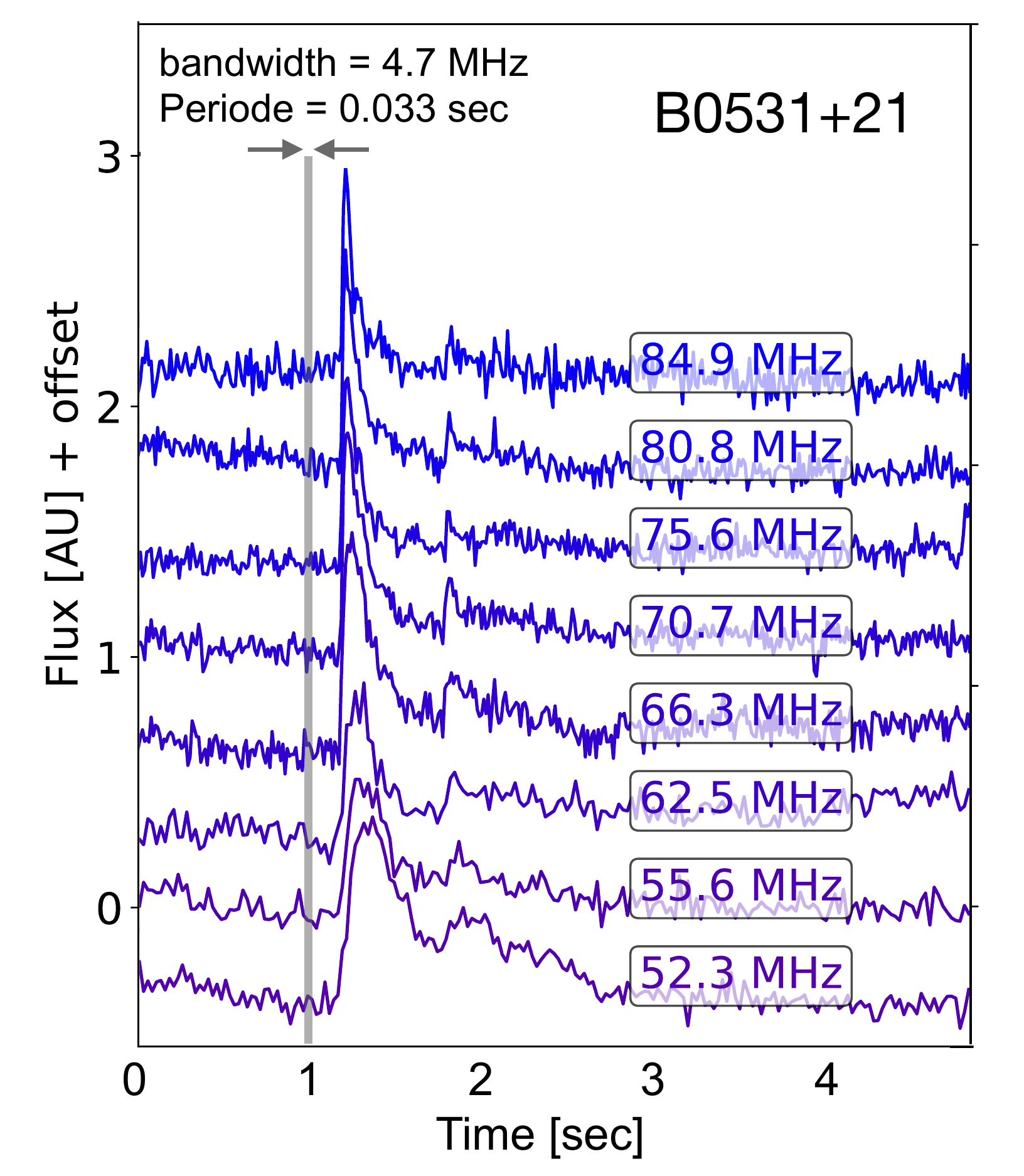}
\caption[Crab pulsar giant pulses]{A giant pulse from the Crab pulsar as a function of time and frequency. The gray area represents a single period of the pulsar.
}
\label{crabe_GP}
\end{figure}

It has long been known that the Crab pulsar emits so-called giant pulses (GPs), very brief bursts of radio emission with intensity many times exceeding the intensity of an average pulse \citep[e.g.~][]{staelin_1968, cordes_crabe_2004, hankins_2016, karuppusamy_2016}. However, at such low frequencies these pulses are highly scattered (at 60\,MHz, the scattering timecale is 5 times the pulsar's spin period) and consequently difficult to observe in folded mode. There have been a few single-pulses campaigns to observe giant pulses at low frequencies: for example, \citet{popov_crabe_2006} reported on the detection of 45 pulses in 12 hours with the Ukrainian UTR-2 radiotelescope below 30\,MHz; \citet{eftekhari_low_2016} 
announced the detection of 1400 pulses in 76 hours of observation with the LWA1. With NenuFAR we detected 600 pulses in only 13.8 minutes at a frequency of 68.7\,MHz (Fig.~\ref{crabe}).
This detection rate and the telescope's sensitivity are very promising for future NenuFAR blind searches for new pulsars in the Northern hemisphere.
A strong individual pulse of this observation as a function of the observed frequency is shown in Fig.~\ref{crabe_GP}. At 52\,MHz the scattering tail of this pulse is roughly 20 times longer than the pulsar spin period.

\subsection{Drifting sub-pulses in PSR~B0809$+$74}

\begin{figure}[!]
\centering
\includegraphics[height=6.5cm, trim=1.5cm 0cm 0cm 0cm]{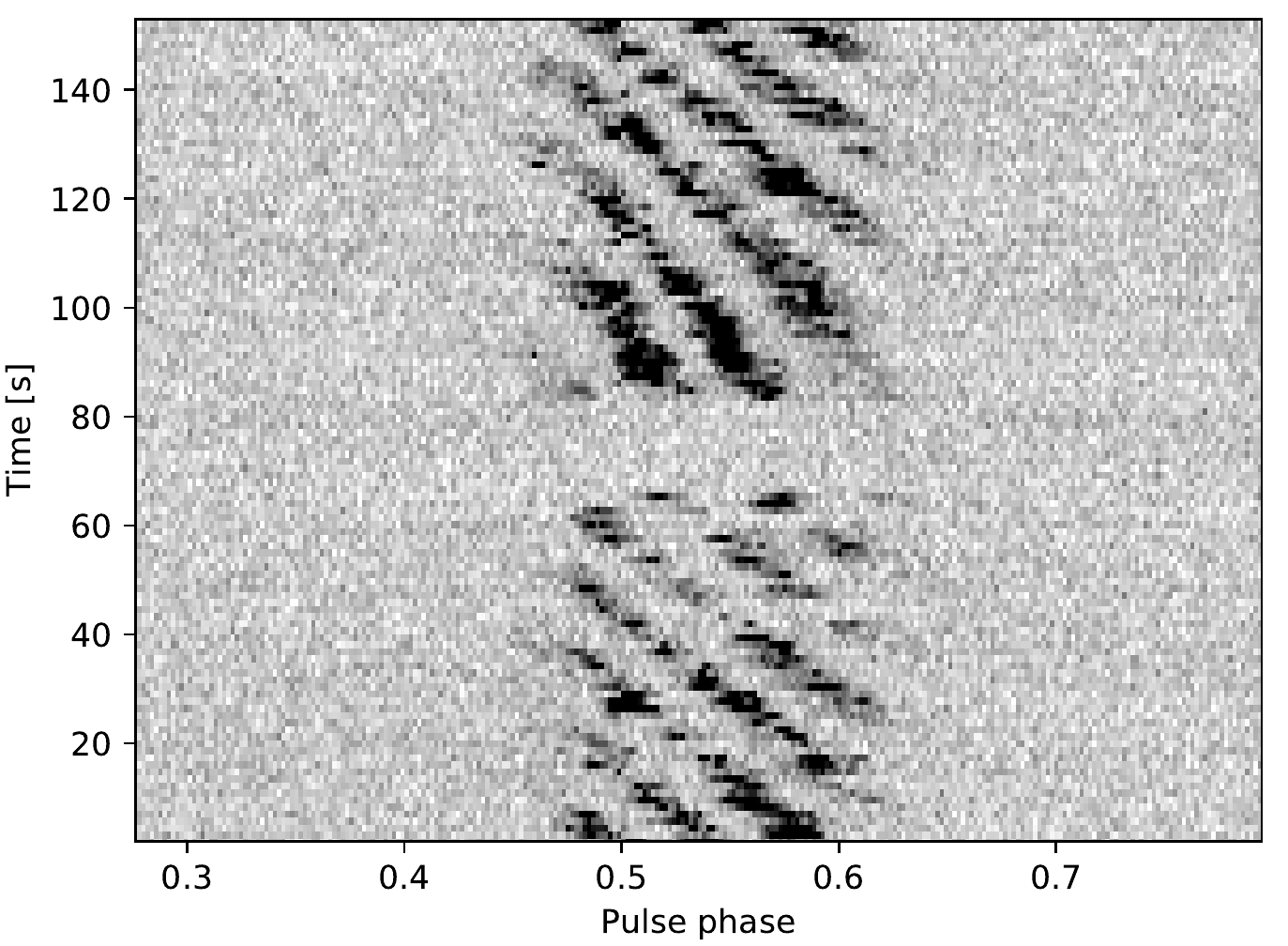}
\caption[B0809+74]{Drifting sub-pulses of PSR~B0809$+$74 in a 4-min observation over a bandwidth of 75\,MHz centered at 47\,MHz. Each line represents a single pulse period. A nulling phase of $\sim$20~s duration is clearly visible
one minute after the start of the observation.}
\label{B0809}
\end{figure}

Drifting patterns are a common phenomenon in pulsars \citep[e.g.~][]{weltevrede_2006, weltevrede_2008, basu_2019}. Many of them show systematic variations of their sub-structures in an organised manner. Among the $\sim$70 known pulsars that exhibit this behaviour, an interesting prototype of this phenomenon is the bright and extensively studied pulsar B0809$+$74. 
It was observed down to 16\,MHz with LOFAR LBA \citep{stappers_observing_2011, hassall_differential_2013}, and detected in the band 18--27\,MHz with UTR-2  \citep{zakharenko_detection_2013}. The drifting sub-structures and the anomalously intensive pulses of PSR~B0809$+$74 \citep[AIPs,][]{ulyanov_2006} are resolved in great details by NenuFAR (Fig.~\ref{B0809}) and are visible all along its frequency domain, from 85 down to 10\,MHz. Because of the large fractional bandwidth and good sensitivity, NenuFAR data are very well suited to study frequency-dependent phase delay of drifting sub-pulses. This delay can provide novel constraints on the location of emission regions in the magnetosphere \citep{bilous_2019, maan_2019}.

\subsection{Detection of 10 MSPs}

\begin{figure}[!]
\centering
\includegraphics[height=7cm, trim=1.5cm 0cm 0cm 0cm]{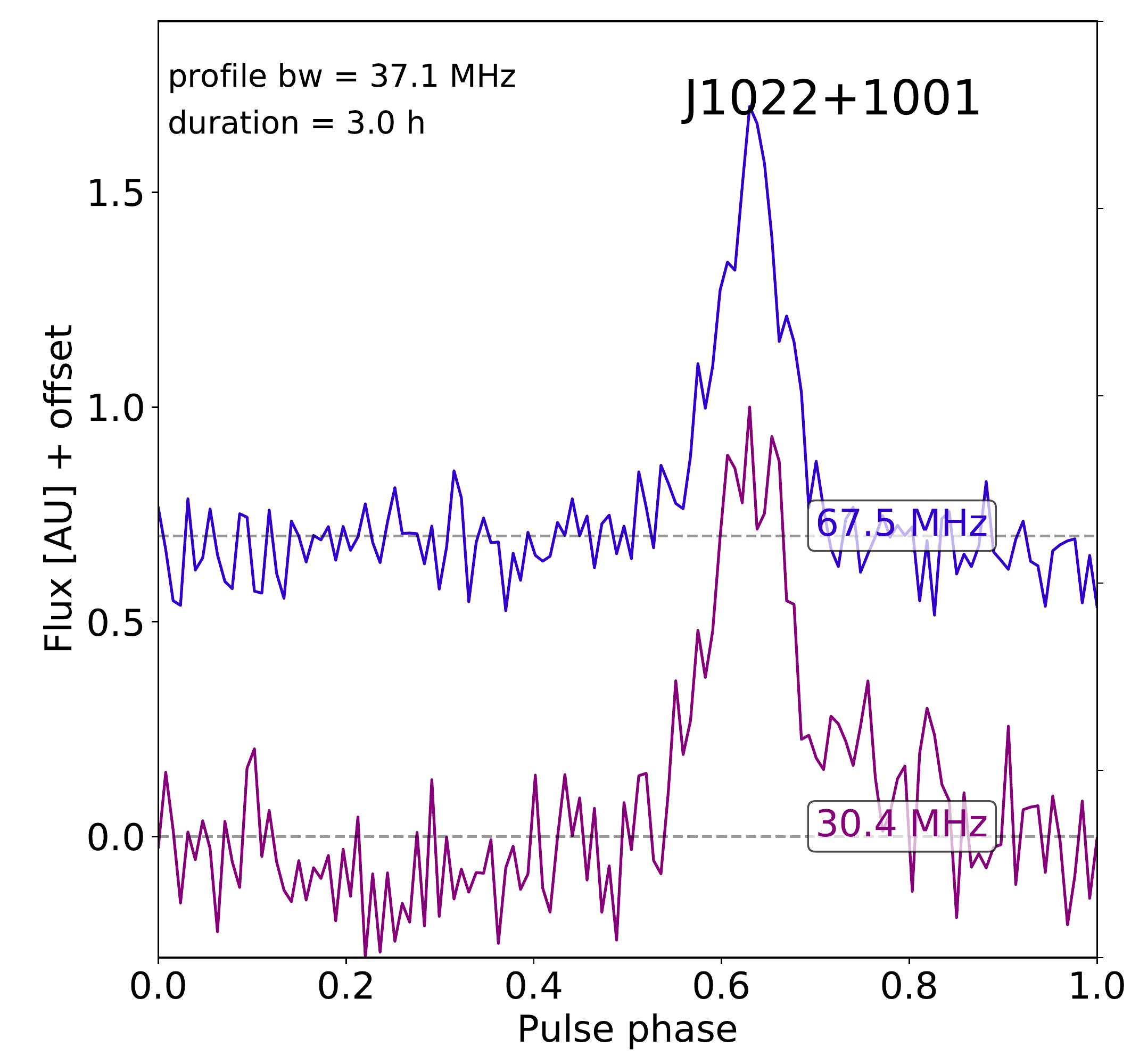}
\caption[J1022+1001]{First detection of the  millisecond pulsar J1022$+$1001 at frequencies below 100\,MHz. 
The observation is integrated over 3 hours and 37.1~MHz of bandwidth at two central frequencies (30.4 and 67.5 MHz) in order to study the frequency dependence of the profile.}
\label{J1022+1001}
\end{figure}

At frequencies below 100\,MHz, only five millisecond pulsars (MSPs) have been detected prior to this work. The first detection of PSR~J2145$-$0750 was published in \citet{dowell_detection_2013}, PSRs J0030+0451 and J0034$-$0534 in \citet{stovall_pulsar_2015}, PSR J1400$-$1431 in \citet{Swiggum_2017} and finally, PSR J0437$-$4715 in \citet{bhat_observations_2018}. The first three MSPs were also reported to be detected with the LOFAR LBA core by \citet{kondratiev_lofar_2016}. Thanks to NenuFAR, we already doubled this list (Bondonneau et al., in prep). As an example, in Fig.~\ref{J1022+1001} we show the observed average pulse profile of PSR J1022$+$1001, obtained during the commissioning phase of the instrument. It was detected with a DM of 10.25356(3)\,pc\,cm$^{-3}$, which is compatible with observations at higher frequencies and the SNR of 46 
corresponding to a mean flux density of $28\pm14$~mJy (following the same method as the one used to calculate the mean flux density limit in section \ref{mean_flux_section}).

\subsection{DM variations from pulsar timing}

The study of time variations in DM for different lines of sight enables studies of the ionised interstellar medium and its turbulence \citep[see, e.g.][]{Phillips91time,Backer93,donner_first_2019}.
This is particularly interesting at low frequencies, where the dispersive time delay is amplified by its dependence on $\nu^{-2}$. 
Such observations demand a large instantaneous fractional bandwidth, high sensitivity, and good clock stability, all of which are provided by NenuFAR.

\begin{figure}
\centering
\includegraphics[height=4.5cm, trim=0cm 0cm 0cm 0cm]{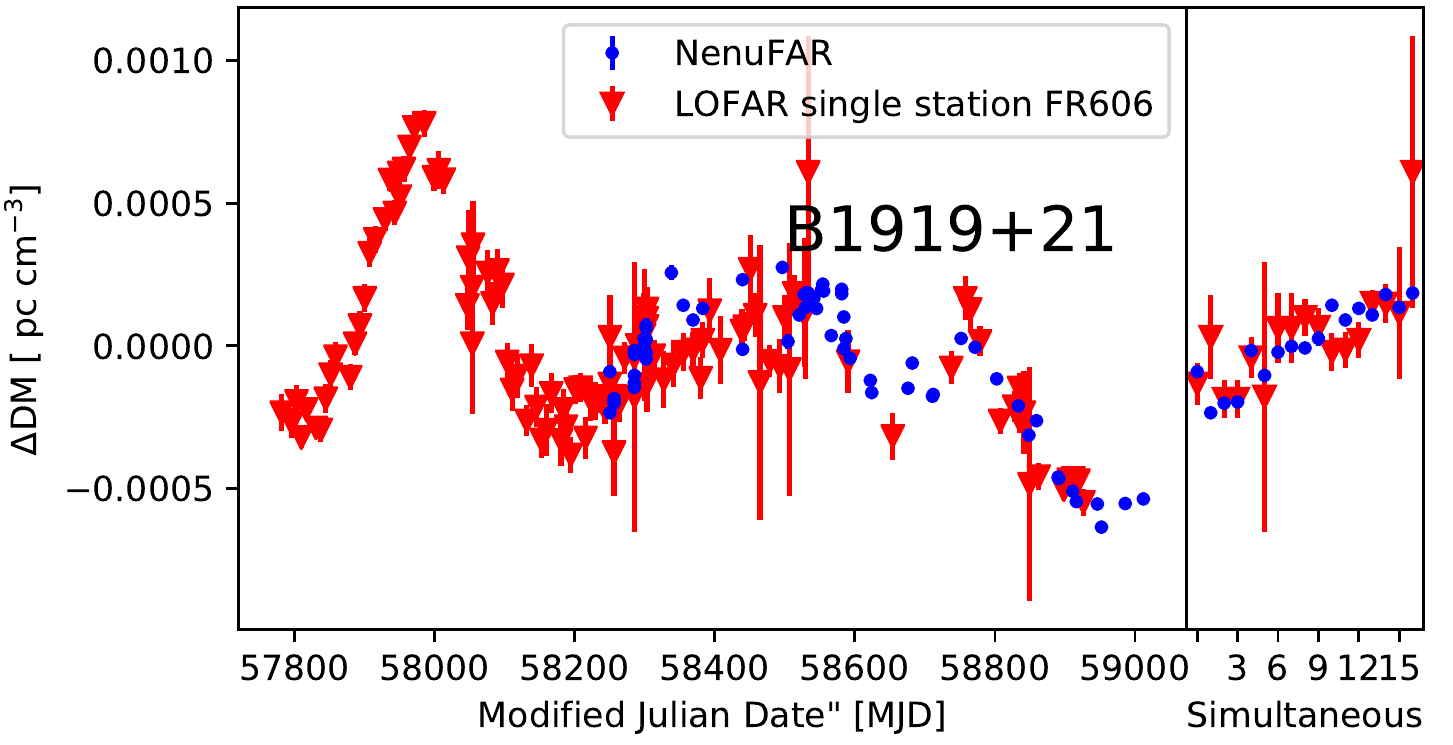}
\caption[B1919+21]
{Right panel: DM variations for PSR B1919$+$21 as measured by NenuFAR (blue dots) and the LOFAR single station FR606 (red triangles) comparatively to the mean value 12.43746\,pc\,cm$^{-3}$ $(\Delta \textrm{DM}=0)$. Right panel: Simultaneous observations between NenuFAR and the LOFAR single station FR606 sorted in chronological order. For NenuFAR measurements the error barre of the DM value is smaller than the point.}
\label{DMofT}
\end{figure}

DM variations are typically studied via pulsar timing and the calculation of pulse times-of-arrival (TOAs).
First tests have shown that for NenuFAR, the typical uncertainties of TOAs are $\sim$5-8 times smaller than for the LOFAR station FR606 (used in LBA mode, i.e.~in the same frequency range). 
This translates to an increased precision for the study of DM variations.
Fig. ~\ref{DMofT} shows DM variations for PSR B1919$+$21 as measured by NenuFAR and LOFAR FR606 obtained with a 2D template fit (a phase-frequency template extending over 118 channels, built from the longest observation of PSR~B1919$+$21 by NenuFAR, showing a SNR > 6000). 
The data from both telescopes are comparable, but the NenuFAR data provide a much higher precision (of the order of $10^{-5}$\,pc\,cm$^{-3}$).
As the template used for pulsar timing is derived from the data itself, the procedure only gives relative DM values (sufficient for the analysis of DM variations) rather than absolute DM values \citep{donner_first_2019}. Series of DMs are well aligned especially if we look at the simultaneous observation in the left panel in Fig. ~\ref{DMofT}. 

\subsection{Interstellar Scintillation Studies}

By recording data in waveform mode, an additional filterbank step can be applied to increase frequency resolution (similar to the dynamic spectrum recording mode) before phase-averaging the observations (as in the folded mode). This hybrid combination of observing modes allows dynamic spectra to be constructed of the pulsed SNR with much higher frequency resolution than the standard folded mode allows; and with more sensitivity than the standard dynamic spectrum mode allows. This is particularly useful for the study of interstellar and interbinary scintillation \citep{ric90}, since the size of scintles is strongly dependent on the observing frequency, causing typical scintillation bandwidths to be much smaller than 195~kHz at NenuFAR frequencies.

\begin{figure}
    \centering
    \includegraphics[width=9.0cm]{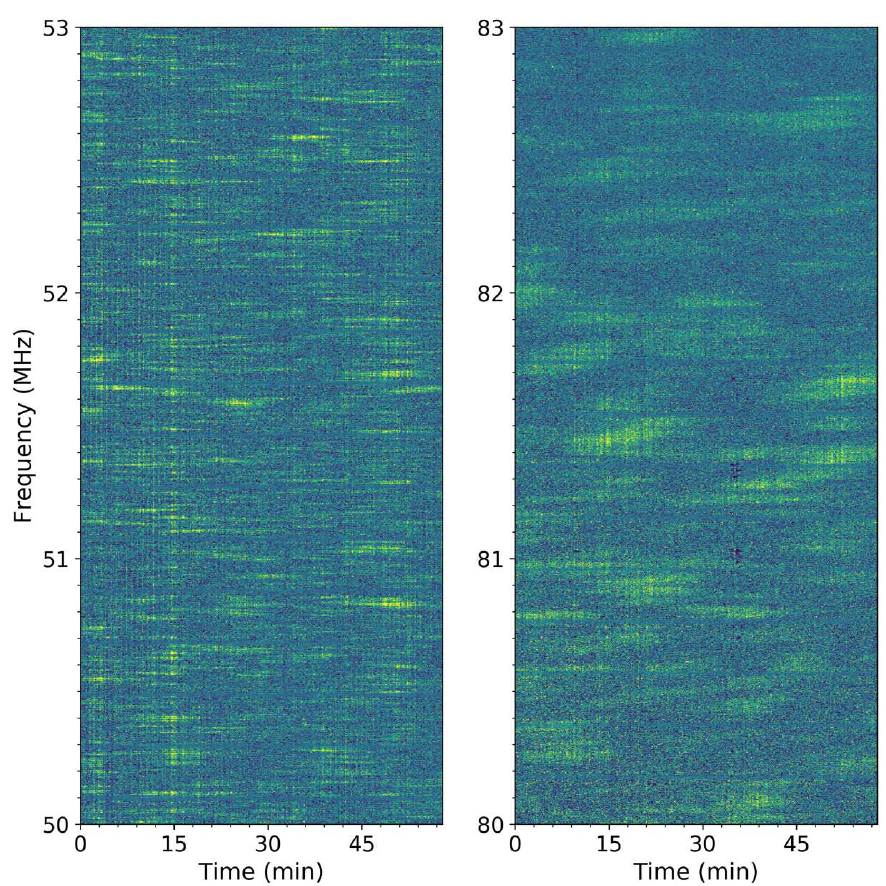}
    \caption{Dynamic spectra of PSR~B0809$+$74, in two 3\,MHz wide bands, of a one-hour NenuFAR observation. The colour scale indicates the pulse SNR ranging from blue (no signal) to yellow (high-SNR), 
    which is heavily modulated due to diffraction in the interstellar medium. The high-SNR "islands" are commonly referred to as scintles and provide information on the turbulent interstellar plasma.}
    \label{fig:DynSpec}
\end{figure}

Figure~\ref{fig:DynSpec} shows an example of a dynamic spectrum that was obtained with this method, on PSR~B0809$+$74. The frequency resolution was set to 1.5~kHz and the data were folded into 10-second integrations. The strong frequency scaling of the scintle size is apparent from the two 3-MHz-wide segments of data shown here: in the 50--53~MHz range the scintillation bandwidth (i.e. the size of scintles in frequency) is $3.9\pm0.7$~kHz; in the 80--83~MHz band it is $23\pm1\,$kHz. The scintillation timescale (the size of the scintles in time) scales less strongly: from $3.79\pm0.07$~min in the 50--53~MHz subband to $6.1\pm0.3$~min in the 80--83~MHz band. This high sensitivity and the fact that even a narrow frequency range can achieve numerous scintles, allows thorough and self-consistent statistical analyses of scintillation scaling laws and enables precise monitoring of scintillation variations and their possible relation to other interstellar effects. 

\subsection{Telescope characterisation}
\subsection{Mean flux density limit}
\label{mean_flux_section}

\begin{figure}
\centering
\includegraphics[height=5.5cm, trim=0cm 0cm 0cm 0cm]{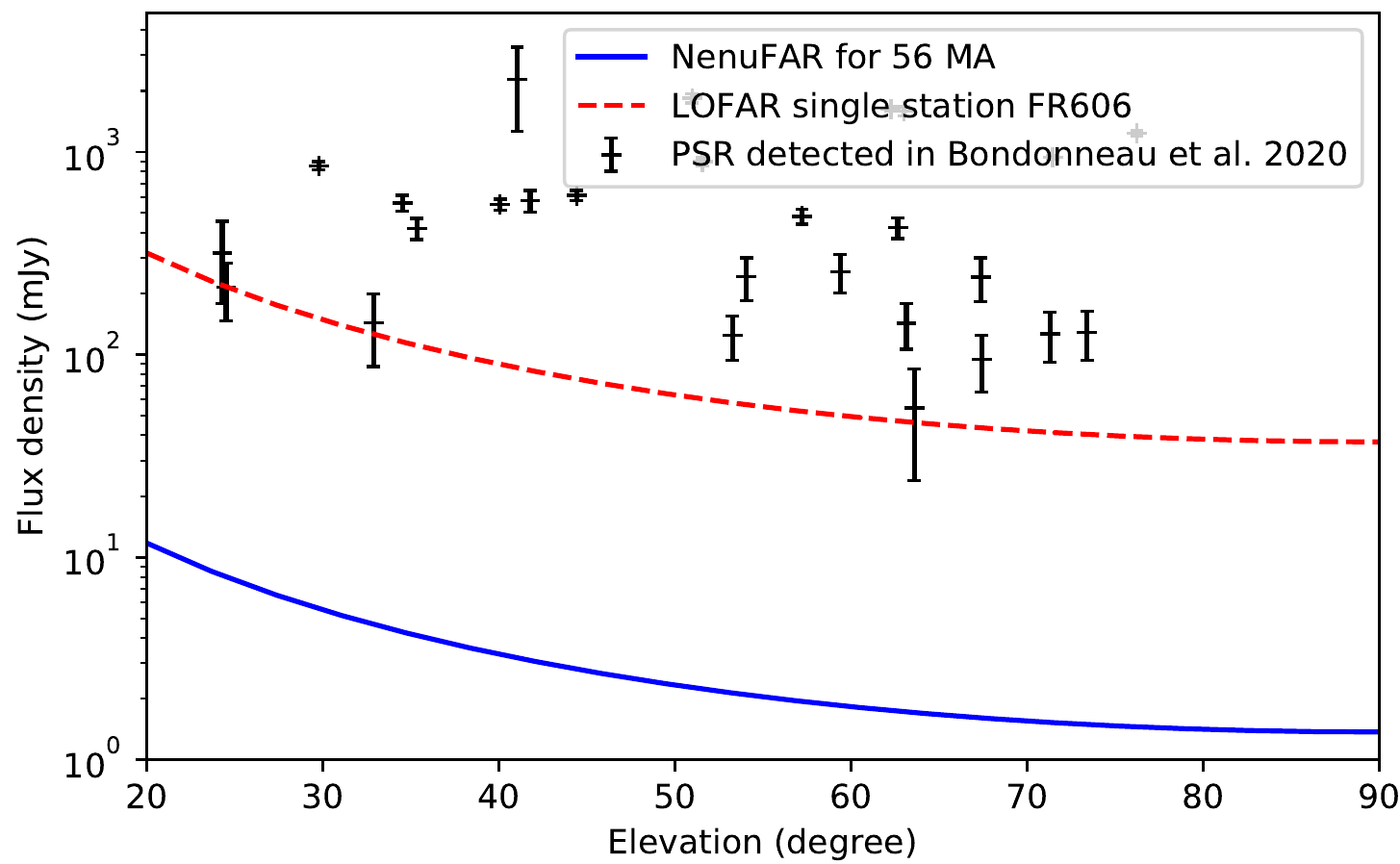}
\caption[mean_flux_elev]{Minimum detectable flux density as a function of source elevation (see text for details). 
The measured mean flux density of the pulsars with duty cycles greater than $>$7\% 
detected with FR606 \citep{bondonneau_2020} are indicated 
in black.}
\label{mean_flux_elev}
\end{figure}

The detection limit of NenuFAR can be estimated using the radiometer equation \citep{lorimer_handbook_2004} with the same set of parameters as in \citet{bondonneau_2020} for the LOFAR single station FR606, but using the collecting area and the instrument temperature of NenuFAR for 56~MA. The result is shown in Fig.~\ref{mean_flux_elev} which compares the minimal flux density detectable by NenuFAR for 56~MA (blue curve) with the corresponding limit 
of FR606 (red curve) and the pulsars detected by FR606. In this Figure the minimum detectable flux density 
is calculated for a pointing outside the Galactic plane ($gl=$\SI{0}{\degree}, $gb=$\SI{90}{\degree}, $T_\text{sky}=2350$ K), assuming 4 hours of integration time, on a 75~MHz bandwidth, a SNR of 5 and a pulse duty cycle of 7\%. The pulsars displayed are detected by the FR606 station with duty cycles greater than 
7\% \citep{bondonneau_2020}. There is a large jump in sensitivity between the instruments, which is mainly due to a factor of 11 in the collecting area. 
This jump will allow for a large number of new detections during the NenuFAR targeted survey.

\subsection{RFI environment at NenuFAR site}
\begin{figure}
\centering
\includegraphics[height=11.0cm, trim=0cm 0cm 0cm 0cm]{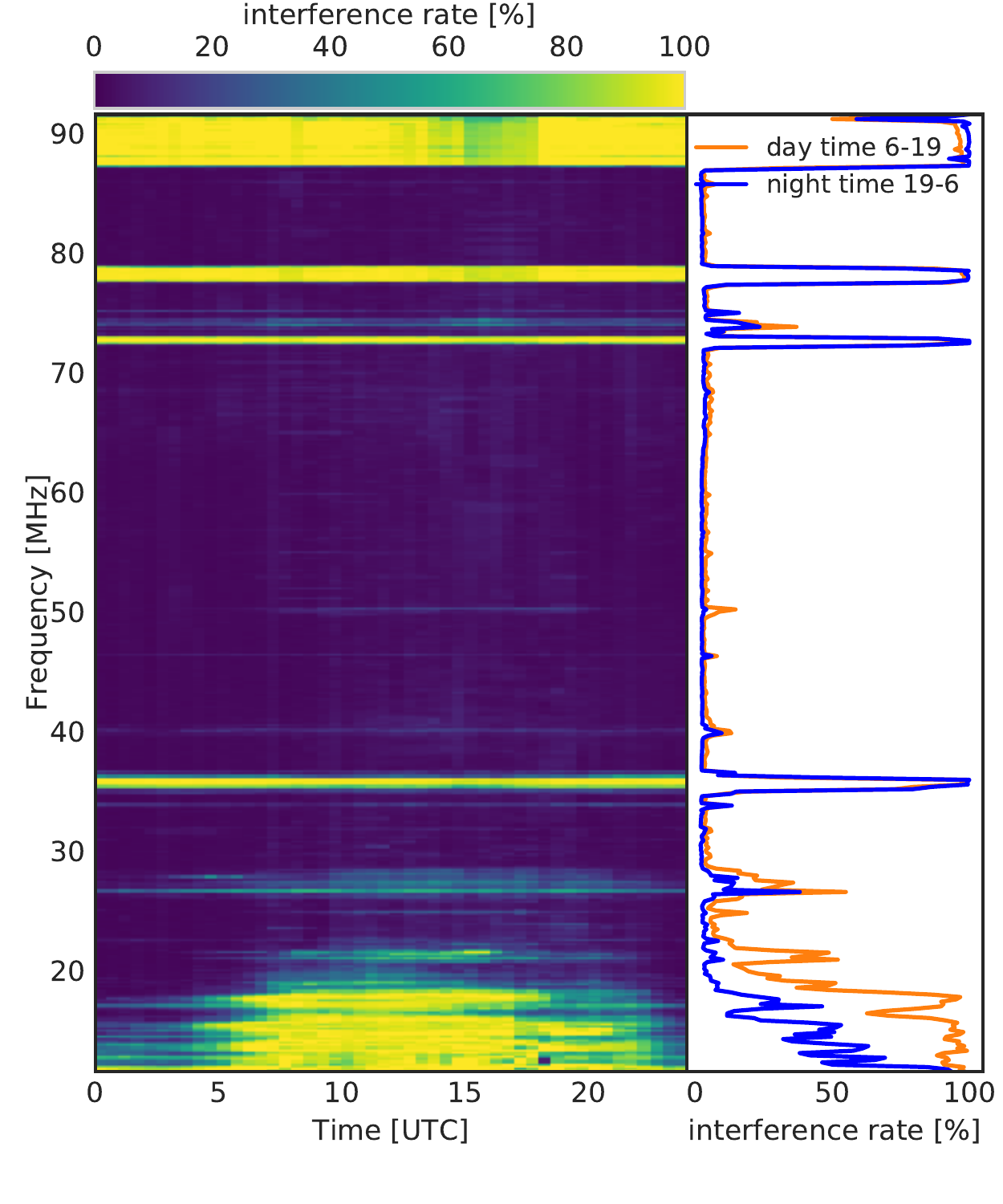}
\caption[RFINenuFAR]
{RFI statistics for NenuFAR computed from more than 1300 hours of observing time (2018--2020). Left panel: Dynamic spectrum of the RFI rate (percentage of data removed during RFI cleaning) as a function of time of day (in UTC) and frequency. Right panel: spectrum of the RFI rate during daytime (orange, 6--19 UT) and night time (blue, 19--6 UT).   }
\label{RFI_NenuFAR}
\end{figure}

Like all radio telescopes NenuFAR is subject to RFI. The post-processing pipeline for the folded mode (Section \ref{section_folded}) is equipped with an automatic RFI cleaner (Section~\ref{RFImitigation}). Based on the observations of the last two years, Fig.~\ref{RFI_NenuFAR} shows RFI statistics across the NenuFAR band. Our observations show a remarkably clean sky between 38 and 72 MHz. A few frequencies are permanently occupied by narrow-band RFI (e.g.~35.5 MHz, 72.5 MHz, and 78\,MHz). Above 87.5\,MHz, the band is saturated by frequency modulation broadcasting. Below 30~MHz Fig.~\ref{RFI_NenuFAR} shows broadband interference (10--25\,MHz) which is much brighter and extends to higher frequencies in the daytime than during nighttime. At daytime, RFI heavily affects the observations of pulsars below 20\,MHz; at nighttime, observations down to 10 MHz remain possible.

\section{Summary}

We have provided a full description of the NenuFAR pulsar instrumentation, including the four main modes used for pulsar observations.
We have shown a number of first results from our observations, 
part of which were obtained already during the commissioning phase. These early science results are very promising and demonstrate the potential of NenuFAR for pulsar science; more detailed follow-up studies are currently being performed within the NenuFAR ``Pulsars Key Science Project''.
We expect significant contributions from NenuFAR, in particular in the fields of pulsar emission mechanisms and the study of the interstellar medium (ISM) and ionosphere.

With its high sensitivity at low frequencies, NenuFAR is also a milestone on the route to the Square Kilometer Array (SKA), hence its pathfinder label\footnote{https://www.skatelescope.org/precursors-pathfinders-design-studies/}.
It will help refining the SKA science goals, particularly concerning emission mechanism and propagation effects in the context of highly dispersed and scattered signals. 
As such, it will contribute to maintain a useful bridge of pulsar studies across the next decade and support the community of pulsar astronomers focusing on emission processes, populations, and ISM characterisation. These science goals are complementary to the ones carried out at L- and S-bands for high precision pulsar timing (gravitational wave detection and tests of General Relativity in the context of the International Pulsar Timing Array collaboration, IPTA). 
When the SKA will be operational, joint studies between the SKA and NenuFAR will be performed in the two hemispheres in complementary frequency ranges, providing the potential for additional discoveries.

\begin{acknowledgements}

This paper is based on data obtained using the NenuFAR radio-telescope. The development of NenuFAR has been supported by personnel and funding from: Station de Radioastronomie de Nan\c{c}ay, CNRS-INSU, Observatoire de Paris-PSL, Université d’Orléans, Observatoire des Sciences de l’Univers en région Centre, Région Centre-Val de Loire, DIM-ACAV and DIM-ACAV+ of Région Ile de France, Agence Nationale de la Recherche.

We acknowledge the use of the Nançay Data Center computing facility (CDN - Centre de Données de Nançay). The CDN is hosted by the Station de Radioastronomie de Nançay in partnership with Observatoire de Paris, Université d'Orléans, OSUC and the CNRS. The CDN is supported by the Region Centre Val de Loire, département du Cher.

This work was supported by the "Entretiens sur les pulsars" funded by Programme National High Energies (PNHE) of CNRS/INSU with INP and IN2P3, co-funded by CEA and CNES.

The Nan\c{c}ay Radio Observatory is operated by the Paris Observatory, associated with the French Centre National de la Recherche Scientifique (CNRS).

\end{acknowledgements}

\bibliographystyle{aa}
\bibliography{Instrumentation_NenuFAR}

%

\end{document}